# The HERMES (High Energy Rapid Modular Ensemble of Satellites) Pathfinder mission

Y. Evangelista[*a,b], F. Fiore[c], R. Campana[d,e], G. Baroni[c], F. Ceraudo[a], G. Della Casa[a], E. Demenev[f], G. Dilillo[a], M. Fiorini[g], G. Ghirlanda[h,i], M. Grassi[j], A. Guzmán[k], P. Hedderman[k], E. J. Marchesini[d], G. Morgante[d], F. Mele[l], L. Nava[h], P. Nogara[m], A. Nuti[a], S. Pliego Caballero[k], I. Rashevskaya[n], F. Russo[m], G. Sottile[m], M. Lavagna[o], A. Colagrossi[o], S. Silvestrini[o], M. Quirino[o], M. Bechini[o], A. Brandonisio[o], F. De Cecio[o], A. Dottori[o], I. Troisi[o], R. Bertacin[p], P. Bellutti[f], G. Bertuccio[l], L. Burderi[m], T. Chen[q], M. Citossi[c], T. Di Salvo[r], M. Feroci[a,b], F. Ficorella[f], N. Gao[q], C. Grappasonni[p], C. Labanti[d], G. La Rosa[m], W. Leone[s], P. Malcovati[j], B. Negri[p], G. Pepponi[f], M. Perri[t], R. Piazzolla[p], A. Picciotto[f], S. Pirrotta[p], S. Puccetti[p], A. Rachevski[u], A. Riggio[v], M. Rinaldi[p], A. Sanna[v], A. Santangelo[k], C. Tenzer[k], A. Tiberia[p], M. Trenti[w], S. Trevisan[c], A. Vacchi[x,u], S. Xiong[q], G. Zampa[u], N. Zampa[x,u], S. Zhang[q], N. Zorzi[f], J. Ripa[z] and N. Werner[z,y]
*on behalf of the HERMES Pathfinder collaboration*

[a]INAF-IAPS Rome, Italy; [b]INFN sez. Roma Tor Vergata, Italy; [c]INAF-OATS, Trieste, Italy; [d]INAF-OAS Bologna, Italy; [e]INFN sez. Bologna, Italy; [f]Fondazione Bruno Kessler Trento, Italy; [g]INAF-IASF Milano, Italy; [h]INAF-OAB, Merate, Italy; [i]INFN sez. Milano Bicocca, Italy; [j]University of Pavia, Italy; [k]IAAT University of Tübingen, Germany; [l]Politecnico di Milano, Como, Italy; [m]INAF-IASF Palermo, Italy; [n]TIFPA-INFN, Trento, Italy; [o]Politecnico di Milano, Milano, Italy; [p]Agenzia Spaziale Italiana, Rome, Italy; [q]IHEP, Beijing, China; [r]University of Palermo, Italy; [s]University of Trento, Italy; [t]INAF-OAR, Rome, Italy; [u]INFN sez. Trieste, Italy; [v]University of Cagliari, Italy; [w]The University of Melbourne, Australia; [x]University of Udine, Italy; [y]Eötvös Loránd University, Budapest, Hungary; [z]Masaryk University, Brno, Czech Republic

## ABSTRACT

HERMES (High Energy Rapid Modular Ensemble of Satellites) Pathfinder is a space-borne mission based on a constellation of six nano-satellites flying in a low-Earth orbit (LEO). The 3U CubeSats, to be launched in early 2025, host miniaturized instruments with a hybrid Silicon Drift Detector/GAGG:Ce scintillator photodetector system, sensitive to X-rays and gamma-rays in a large energy band. HERMES will operate in conjunction with Australian Space Industry Responsive Intelligent Thermal (SpIRIT) 6U CubeSat, launched in December 2023. HERMES will probe the temporal emission of bright high-energy transients such as Gamma-Ray Bursts (GRBs), ensuring a fast transient localization in a field of view of several steradians exploiting the triangulation technique. HERMES intrinsically modular transient monitoring experiment represents a keystone capability to complement the next generation of gravitational wave experiments. In this paper we outline the scientific case, development and programmatic status of the mission

**Keywords:** High Energy Astrophysics, HERMES, SpIRIT, CubeSat, Payload, Space 4.0, Silicon Drift Detectors, GAGG, ASIC, Gamma-ray Bursts

## 1. INTRODUCTION

The most dramatic events in the Universe, the death of stars and the coalescence of compact objects to form a new black hole, produce among the most luminous objects: Gamma Ray Bursts (GRB). However, most of the light is produced quite far from where the action is, i.e., the newborn event horizon, the accretion disk, and the region from which a relativistic jet is launched. On the other hand, gravitational waves (GWs), encoding the rapid/relativistic motion of compact objects, give us a direct look into the innermost regions of these systems, providing precise information on space-time dynamics such as mass, spin, inclination, and distance. This information can be greatly enhanced by identifying the context in which the event occurs, which can be done via electromagnetic observations, as the GW/GRB170817 event has strikingly shown [1].

---

[*]Send correspondence to Y. Evangelista: yuri.evangelista@inaf.it; https://www.hermes-sp.eu

In fact, the GW170817 event marks the beginning of the so-called multi-messenger astrophysics, in which new observations of Gravitational Waves (GW) added up to traditional electromagnetic observations from the very same astrophysical source. The nearly simultaneous detection of a GW signal from merging Neutron Stars (NS) by the Advanced LIGO/Virgo and of a short GRB detected by the Fermi and INTEGRAL satellites [2], proved that short GRBs are due to merging NSs (or NS-BH), as hypothesized since about thirty years. Furthermore, the quick localization of the GW-GRB event produced the discovery and the detailed study of a kilonova event. Advance in technology of the GW detectors will significantly increase the discovery volumes making more challenging to spot electromagnetic counterparts. Fortunately, the number of X-ray transients in the same volume is much more limited (from zero to a few), making much more efficient the search for the electromagnetic counterparts of GW events in this energy band.

The operation of an efficient X-ray/γ-ray all-sky monitor with good localization capabilities will have a pivotal role in bringing multi-messenger astrophysics to maturity and will fully exploit the huge advantages provided by adding a further dimension to our capability to investigate cosmic sources. The HERMES (*High Energy Rapid Modular Ensemble of Satellites*) Pathfinder program offers a fast-track and affordable complement to more complex and ambitious missions for relatively bright events.

HERMES [3] is a mission concept based on a constellation of 6 nano-satellites (3U CubeSats) in low Earth orbit. The constellation is built upon a twin project: the HERMES Technological Pathfinder (HERMES-TP), funded by the Italian Ministry for Education, University and Research and the Italian Space Agency, and the HERMES Scientific Pathfinder (HERMES-SP), funded by the European Union's Horizon 2020 Research and Innovation Programme under Grant Agreement No. 821896. Both projects (HERMES-TP and HERMES-SP) provide three complete satellites to the constellation (i.e. payloads, PL, and service modules, SVM) aiming at demonstrating that fast GRB detection and localization is feasible with disruptive technologies on-board miniaturized spacecrafts, mostly exploiting commercial off-the shelf (COTS) components. Moreover, the Italian Space Agency approved and funded the participation to the SpIRIT (*Space Industry – Responsive – Intelligent – Thermal Nanosatellite*) CubeSat [4]. The SpIRIT project, which is supported by the Australian Space Agency and led by University of Melbourne, hosts a HERMES-like detector, thus providing a seventh unit to the HERMES constellation. SpIRIT is a 6U CubeSat and has been launched in a Sun-synchronous Orbit (SSO) on December 1$^{st}$, 2023. It is currently in a commissioning phase and scientific observations are expected to start in the second half of 2024 [5].

At the core of the HERMES Pathfinder mission is a hybrid detector concept that is capable of measuring both soft X-rays as well as γ-rays. The selected design consists of GAGG:Ce scintillator crystals optically coupled to Silicon Drift Detectors (SDDs). The SDDs can directly detect the soft X-rays up to ~30 keV, while for higher energies the SDDs collect the light produced by the γ-rays in the scintillator crystals, effectively extending the detector sensitivity to the MeV range. The transient event position is obtained by estimating the delay between the arrival time of the signal on different detectors, placed hundreds/thousands of kilometers away. The accuracy in determining the position depends on several aspects, such as the cross-correlation calculated between light curves obtained from the observation of the same event from different detectors; the uncertainty on absolute time of each photon collected; the error due to the uncertainty on the location of the satellite; possible systematic uncertainties on the method; the average baseline, namely the average distance between detectors and the number of detectors which see a given event simultaneously. Once the position of the transient is determined and after correcting for the delay time of arrival, the signals received by the different detectors can be combined. This operation will significantly increase the collecting area and thus the sensitivity to finer temporal structure.

## 2. HERMES SCIENTIFIC AND MISSION REQUIREMENTS

The overall main scientific goal of the HERMES project is to provide a system able to detect and accurately localize GRBs and other high-energy transients, such as the counterparts of GW events (merging of compact objects, supernovae, SN or other unexpected phenomena), that can be deployed in a few year, bridging the gap between the aging, past generation of X-ray monitors (Swift, INTEGRAL, AGILE and Fermi) and the next ones. This scientific goal translates into the two main mission requirements of the HERMES project:

**MIS-REQ-1:** Detect GRBs with peak flux ≥ 0.5–1 ph/s/cm$^2$ in the 50–300 keV band.

**MIS-REQ-2:** Detect ≥ 40 long GRBs and ≥ 8 short GRBs simultaneously in at least 3 units with an efficiency ≥ 40–50% in each unit, to be able to assess their position through the analysis of the delay time in the signal collected on different detectors.

Moreover, the fulfilment of the above-mentioned mission requirements will allow HERMES Pathfinder to:

a) validate the overall concept; study the statistical and systematic uncertainty on both detection and localization to validate/improve the payload and service module design and make it more reliable for the proposal of a large constellation.

b) prove that accurate timing in the still little explored window between a fraction of μs to 1 ms is feasible using detectors with relatively small collecting area. (It is worth noticing that, although hosted in a nanosatellite platform, the HERMES timing resolution is ~300 ns, about 7 times better than Fermi/GBM).

c) study the uncertainties associated with the addition of signals from different detectors to improve the statistics on high resolution time series.

A detailed study of HERMES scientific performance can be found in [6] and [7], while the temporal techniques that allow HERMES localization and all-sky monitoring capabilities are discussed in [8].

## 3. HERMES PAYLOAD DESIGN AND ANTICIPATED PERFORMANCE

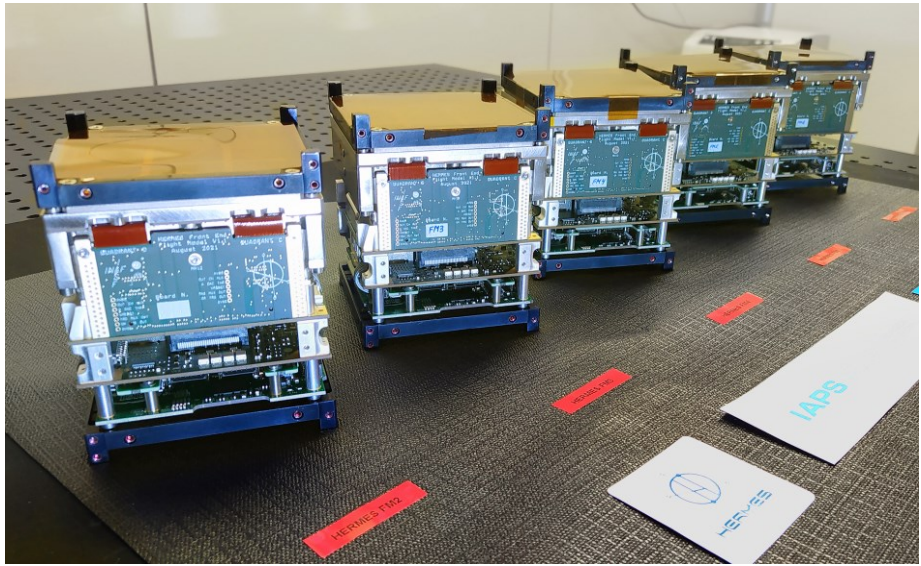

**Figure 1** Five out the six HERMES Pathfinder payloads (H2 to H6) integrated and tested in the INAF-IAPS Rome clean room. When this photograph was taken, H1 payload and SpIRIT/HERMES payload were already integrated in the respective spacecraft.

The HERMES scientific and mission requirements flow-down to ambitious payload requirements, including broad energy band, high detection efficiency, good energy resolution, sub-microsecond temporal resolution, compact and lightweight design, reliable operation in a quite broad range of space environments (e.g., temperature, radiation damage, etc.).

To fulfill the broad energy range requirement, an integrated detector [9][10][11] has been developed to exploit a "double detection" mechanism, with a partial overlap of the two different detection systems efficiency around ~30 keV. Detection of soft X-rays (hereafter *X-mode*) is obtained by a segmented solid-state detector employing custom designed Silicon Drift Detectors (SDD), with a cell size of about 0.45 cm$^2$, which allows to attain a low noise level (of the order of a few tens of e$^-$ rms at room temperature) and correspondingly a low energy threshold for the detection of X-ray radiation (~3 keV). On the other hand, detection of hard X-rays and γ-rays (*S-mode*) exploits the conversion of high-energy photons into visible light by means of scintillator crystals. Scintillation-generated optical photons are then collected and converted into electric charge in the same photodetector (SDD) used to directly detect X-ray photons. In this case, the

SDDs act as a photodiode and produce an amplitude charge signal proportional to the amount of scintillation light collected.

The discrimination between the two signals in the SDD (soft X-rays or optical photons) is achieved by means of the segmented detector design. Each scintillator crystal is read-out by two SDD cells, so that events detected in only one SDD are associated with soft x-rays converted in a single SDD cell, while events detected simultaneously in more than one SDD are generated by the optical light produced in the scintillator by an incoming hard X-ray/γ-ray. The low-noise signal read-out is ensured by state-of-the-art, CMOS ASICs (LYRA-FE and LYRA-BE [12]), specifically developed for the HERMES Pathfinder mission.

### 3.1 Payload anticipated performance

Table 1 summarizes the main HERMES Pathfinder payload performance, while Figure 2 shows the PL effective area as a function of the boresight angle and for an azimuthal angle φ = 90°, obtained through a Monte Carlo simulation of a detailed mass model of the payload and of the satellite platform (the off-axis effective area for different azimuthal angles is slightly different due to the variations in absorption/scattering by passive materials surrounding the detector).

**Table 1** HERMES Pathfinder payload anticipated performance

| PARAMETER | VALUE |
| --- | --- |
| Payload peak effective area (X-mode & S-mode) | 52 cm$^2$ |
| Field of View | 3.2 sr FWHM |
| Low energy threshold | ≤ 3 keV |
| Energy resolution X-mode (@ 6 keV) | ≤ 800 eV FWHM |
| Energy resolution S-mode (@ 60 keV) | ≤ 5 keV FWHM |
| Time resolution (68% c.l.) | 320 ns |
| Time accuracy (68% c.l.) | 181 ns |
| Background rate 50–300 keV | 72 counts/s |
| Background rate 3–2000 keV | 510 counts/s |
| Payload mass | 1.55 kg |
| Payload power in Observation | 1.8 W |
| Payload Telemetry (scientific + engineering) | 610 Mbits/day (50% duty cycle) |

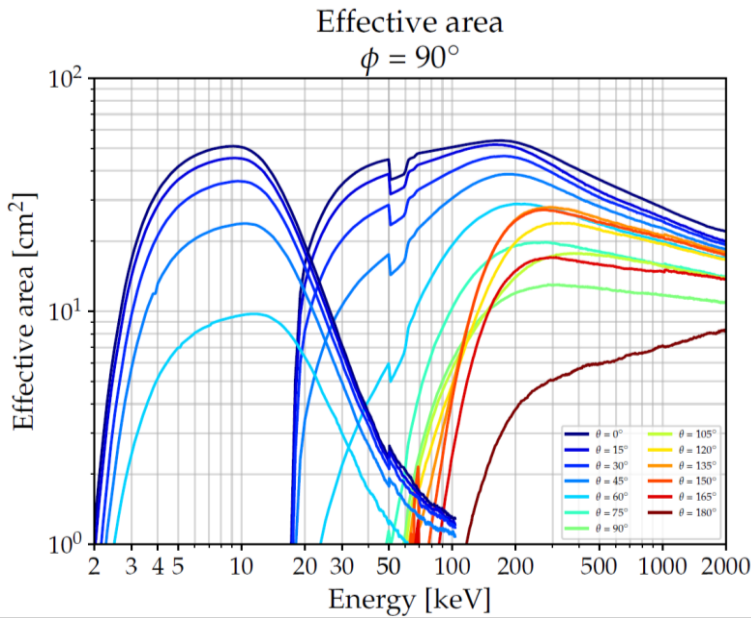

**Figure 2** HERMES Pathfinder PL effective area as a function of the boresight angle and for an azimuthal angle φ = 90° (the off-axis effective area for different azimuthal angles is slightly different due to the variations in absorption/scattering by passive materials surrounding the detector)

As reported in [6], the GRB detection rate for the HERMES Pathfinder constellation has been estimated considering a duty cycle of 50% (expected in a SSO orbit) and that the six satellites are arranged in two triplets observing different regions of the sky. In this configuration, each satellite triplet covers a field of view of ~5.20 sr (*S-mode*) and ~3.14 sr (*X-mode*), leading to the rates reported in Table 2.

Table 2 Long and short GRB detected events per year expected with HERMES (adapted from [6]). For each GRB class, the table reports the total rate and the partial rates for the S and X modes (S-mode: events detected in S-mode; X-mode: events detected in X-mode; X-only: events detected only in the X-mode)

| GRB CLASS | S-MODE | X-MODE | X-MODE ONLY | TOTAL |
|---|---|---|---|---|
| Long | 147 | 131 | 51 | 195 |
| Short | 19 | 0.3 | <0.1 | 19 |

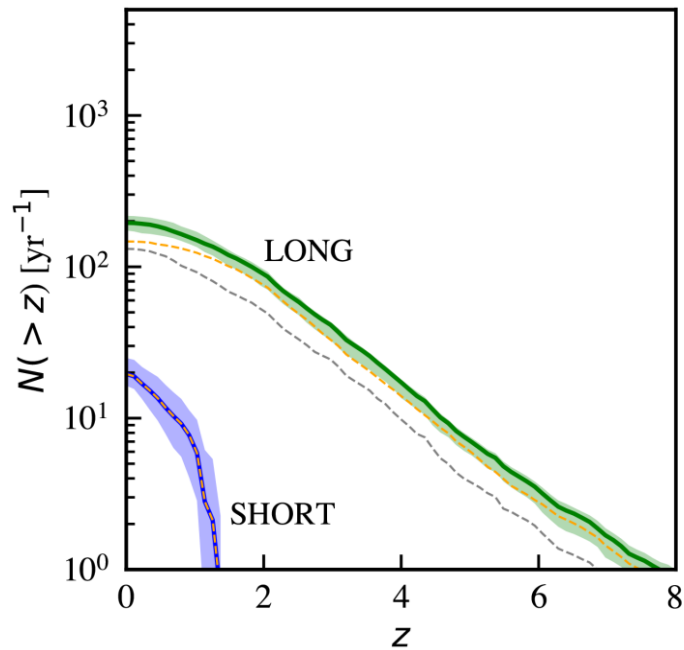

Figure 3 Expected HERMES Pathfinder GRB detection rates (events yr$^{-1}$) as a function of redshift $z$

Considering the sample of 8 high-redshift ($z > 6$) GRBs observed over the last two decades, it is possible to estimate that, thanks to the energy band extension down to soft X–rays (3 keV), HERMES Pathfinder will be able to detect ~3.4 yr$^{-1}$ GRBs at redshift $z > 6$ (Figure 3). Thus, in presence of NIR follow-ups provided by ground based and/or spaceborne facilities (also on-board nanosatellites, e.g., SkyHopper [13]; CHIPS [14]), HERMES Pathfinder will be able to significantly increase the high-$z$ GRB sample already during its operative life.

## 3.2 Payload design

Figure 1 shows five out the six HERMES Pathfinder payloads (H2 to H6) integrated and tested in the INAF-IAPS Rome clean room, while in Figure 4 an exploded view of the HERMES Pathfinder payload is reported. The payload is composed of 4 main subsystems: the detector assembly (DA), the back-end electronic board (BEE), the power-supply electronic board (PSU) and the payload data handling unit (PDHU).

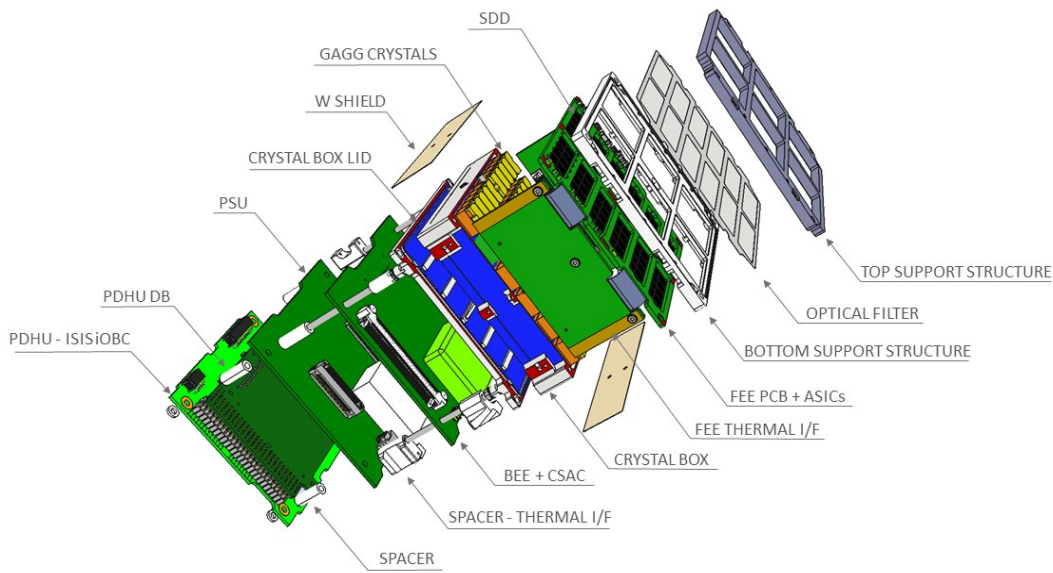

**Figure 4** Exploded view of the HERMES Pathfinder payload

The HERMES detector assembly is composed of:

**Optical filter and MLI**, manufactured by the IHEP of Beijing (China), the filter is made of 300 nm aluminum deposited on a thin (1 μm) polyimide foil. The optical filter primary task is to prevent O/UV light from reaching the SDD detector, minimizing the current noise generated in the NIR/O/UV band. Moreover, being the HERMES PL thermal design based on passive cooling only, the filter also contributes to the overall thermal design of the detector assembly. An additional multilayer insulation film (MLI) is mounted on top of the detector after the final mechanical assembly.

**Front-end electronics** (**FEE**) board (Figure 5), a rigid-flex printed circuit board with a slotted main plate and two side wings, connected to the main part with a flexible flat cable integrated in the board.

The FEE PCB hosts:

- 12 Silicon Drift Detector (SDD) arrays (Figure 6), each one with 10 independent cells, for a total of 120 detector channels.
- 120 LYRA-FE ASIC dies, one for each SDD channel.
- 4 LYRA-BE ASICs, each one interfaced with 30 LYRA-FE channels.

**Scintillator crystals** (Figure 7), each one optically connected to two SDD cells. To maximize the optical contact between SDD and crystal, ensuring optimal scintillation light readout, each crystal is:

- optically isolated on each side except the one in direct contact with the SDD.
- optically coupled to the SDD through a transparent silicone layer.

**Crystal box,** hosting the 60 scintillator crystals and providing the overall DA mechanical structure together with the FEE board and the detector support structure. Three 200 μm thick tungsten sheets are glued on the bottom side and on two lateral sides of the crystal box to provide background shielding.

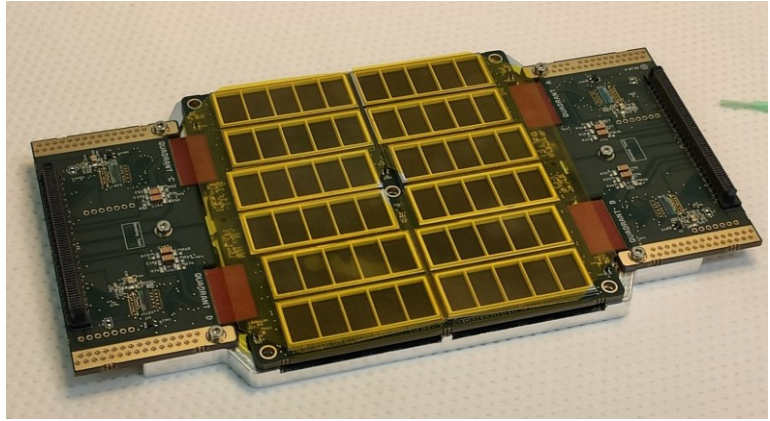

**Figure 5** Front-end PCB (FEE), integrated with 12 SDD arrays and LYRA-FE and LYRA-BE ASICs

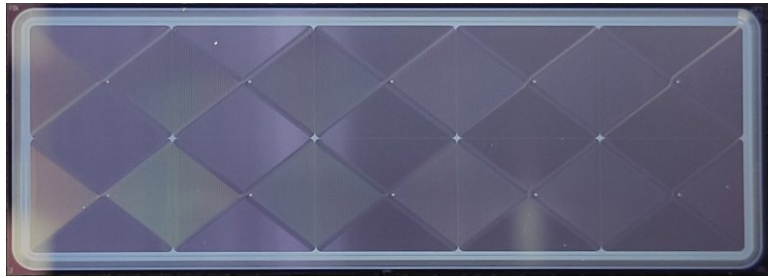

**Figure 6** 2×5 SDD matrix (n-side/anode side). Each SDD cell is 7.44×6.05 mm$^2$

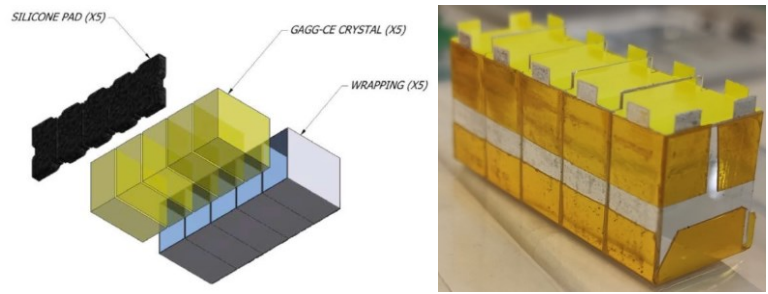

**Figure 7** Ce:GAGG crystal optical assembly

The BEE (Figure 8) is the logic block between the front-end ASICs and the Payload Data Handling Unit. The BEE oversees the ASICs configuration, the analog to digital conversion of the ASIC signals and the event time-tagging exploiting the sub-microsecond accuracy of a local chip-scale atomic clock (CSAC). Moreover, the BEE collects the detector analog and digital house-keeping data, commands the power lines required by the FEE, manages events data acquisition, transmits science data and HKs to the PDHU as well as to receive and decode tele-commands sent by the PDHU. The core of the BEE is a SEL immune Intel/Altera Cyclone V FPGA that implements all the required functions and tasks.

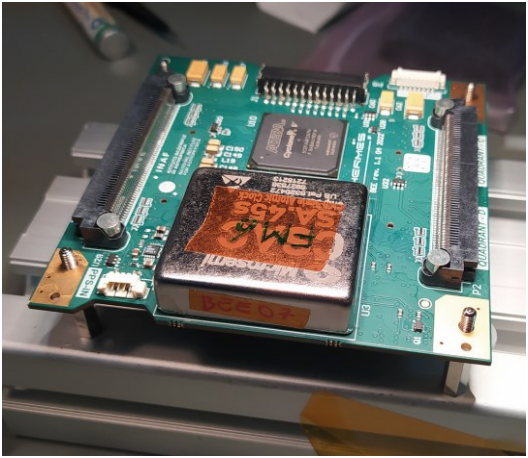
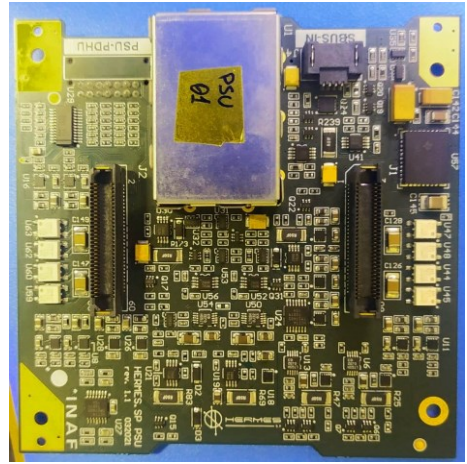

**Figure 8** Back-End Electronics (BEE) PCB

**Figure 9** Power Supply Unit (PSU)

A custom Power Supply Unit (PSU, Figure 9) board has been designed and manufactured to provide the power supplies required by the payload [15]. All the low voltages needed for the operation of the FEE are generated by ultra-low noise linear Low Drop-Out (LDO) regulators. A specialized DC-DC converter is used for the generation of the high-voltage detector biases. Low voltage lines are protected against overcurrent or latch-up in the powered circuitry by means of current monitors which also provide alert lines routed to the BEE FPGA for latch-up monitoring and logging. Special care has been taken in the PSU design to ensure a robust and safe latch-up control. A delay circuit is used to avoid the protection activation due to inrush current, and two of sense amplifiers are implemented in parallel in a redundant configuration for high reliability protection of sensitive power lines.

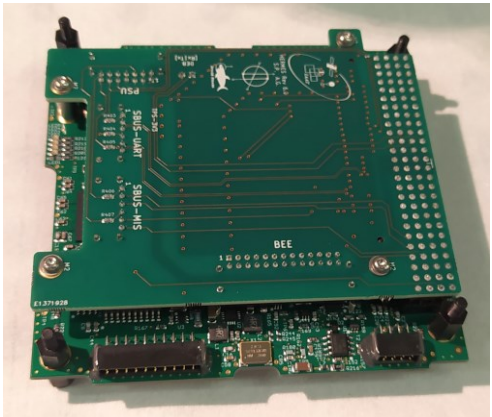
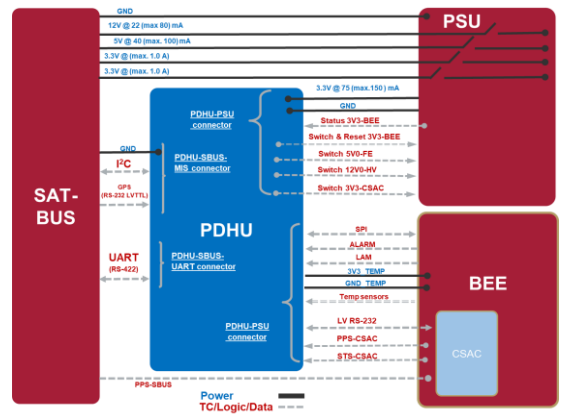

**Figure 10** Left: Payload Data Handling Unit (PDHU) in the motherboard-daughterboard configuration. Right: Schematic view of the P/L electrical interfaces

The Payload Data Handling Unit (PDHU, Figure 10) is the interface between the spacecraft and the payload. The selected hardware for the PDHU is the Innovative Solutions In Space (ISIS) On-Board Computer (iOBC). The iOBC is a flight proven, high performance processing unit based around an ARM9 processor. Combined with a custom-made daughterboard, the PDHU provides all the payload-bus electrical interfaces, the payload central processing unit (CPU) and mass memory, and it oversees interfacing the PSU for voltage line commanding, the BEE for internal TM/TC transmission and atomic clock configuration, and the analog temperature sensors of the payload for temperature monitoring and logging. Moreover, the PDHU manages the payload operative modes, generates and filters the photon list, provides the formatting of the scientific and housekeeping data and performs the burst trigger search. A detailed description of the PDHU and of its functionalities and performance can be found in [16][17].

## 4. HERMES SERVICE MODULE DESIGN

The HERMES Pathfinder service module has been designed and developed by Politecnico di Milano to fulfill the mission requirements and scientific goals reported in Section 2. In particular, the SVM shall be able to:

- Provide absolute position of each S/C with an accuracy ≤30 m (3σ)
- Provide synchronization information about coordinated timing to the PL (e.g. GPS/PPS signal)
- Provide photon-by-photon scientific observations data within 24 hours from their collection
- Distribute GRB trigger messages to ground within 30 minutes
- Provide 3 DOF pointing stability with an accuracy better than 5°
- Provide power to the PL and other subsystems for the whole mission duration
- Withstand LEO radiation environment
- Provide the thermal environment (temperature ranges, gradients, and stability) required to ensure full performance during each mission phase

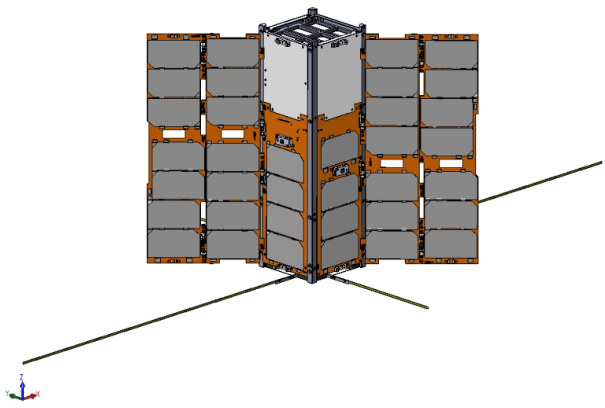

**Figure 11** HERMES spacecraft front view

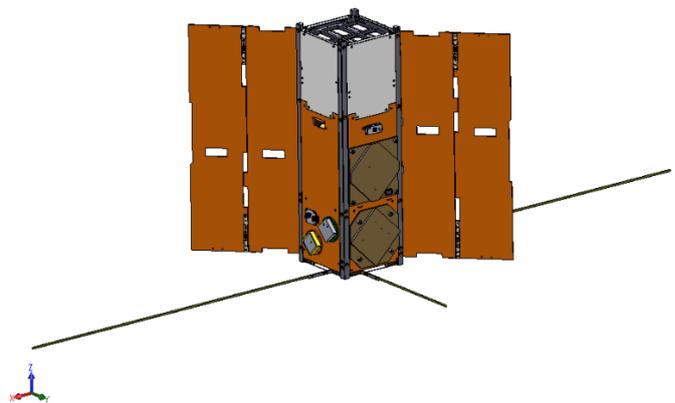

**Figure 12** HERMES spacecraft back view

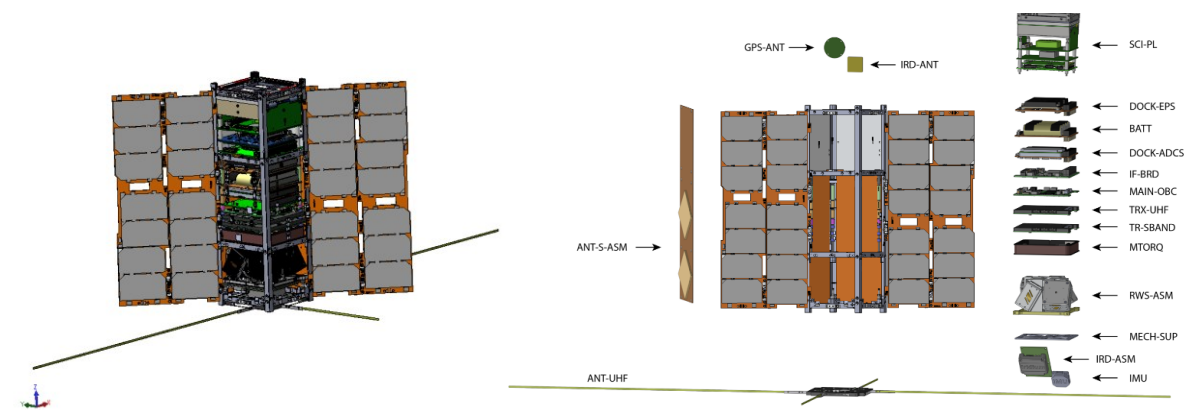

**Figure 13** HERMES open front view (left) and exploded view (right)

A global overview of the HERMES Pathfinder CubeSat configuration is illustrated in Figure 11 and Figure 12, where the front view (Sun) and the back view (Deep Space) are represented. It is composed of the main 3U body, on which are mounted deployable solar arrays and UHF/VHF antenna (bottomside, –Z). The bottom and the middle U are dedicated to the S-band antennas (RX/TX) on the +Y face, while the +X bottom U face is dedicated to GPS and Iridium antennas. The top U, which contains the payload, has all faces left free from solar cells and coated in white paint for thermal constraints.

The HERMES Pathfinder satellite internal configuration is shown in Figure 13 and it is organized as follows:

- Top Unit: fully dedicated to the payload
- Central Unit: dedicated to batteries, control boards and all the electronics.
- Bottom Unit: mostly dedicated to reaction wheels

Table 3 and Table 4 summarize the stack components and the lateral faces components, while Figure 14 shows the HERMES H2 flight model during assembly, integration, verification and test (AIV/T) activities in Politecnico di Milano ISO 8 clean-room. Further information about the HERMES Pathfinder SVM and mission planning can be found in [18][19][20][21].

Table 3 Satellite stacked component description

| ACRONYM | FULL NAME | NOTES |
|---|---|---|
| PL | Scientific Payload | Fits the all the top U, harness and electronics included |
| BATT | Battery Pack | |
| DOCK-EPS | EPS docking board | Fits the power distribution and control units: Power Distribution unit (PDU) and Array Conditioning Unit (ACU) |
| DOCK-ADCS | ADCS Docking Board | Fits a dedicated OBC for attitude control (ADCS-OBC) and the GPS board. |
| IF-BRD | Interface board | Fits data and power connections to interface COTS components. |
| MAIN-OBC | OBC Board | |
| TRX-UHF | UHF Transceiver | |
| TR-SBAND | S-Band Transmitter | |
| MTORQ | Magnetorquers | Mounted on the bottom side of the PC104 stack |
| RWL-ASM | Reaction-Wheels | Mounted in the bottom U. Dedicated standoffs are used for mounting. |
| MECH-SUP | Mechanical Support | Assembly with dedicated fixing on bottom of the CubeSat structure and consist of a mounting board (STR-SUP1) on which are mounted: the Iridium Assembly Board (IRD-ASM), the IRIDIUM modem (IRD-MDM) and interface board (IRD-IF), and an Inertial Measurement Unit (IMU). |
| UHF-ANT | UHF Antenna | Mounted on the bottom side of the structure to avoid interference with the DSA. This position allows for deployment even in case of solar array deployment failure. |

Table 4 Lateral faces component description

| ACRONYM | FULL NAME | NOTES |
|---|---|---|
| GPS-ANT | GPS antenna | |
| IRD-ANT | IRIDIUM antenna | |
| ANT-S-ASM | S-band antenna assembly | Consist of the two patch antennas (RX an TX) integrated on the middle and bottom +Y face. |
| FSS | Fine sun sensors | Mounted on ±X and Y direction with an inclination of 6° respect to the panel plane, to avoid FoV interference by DSA once deployed. |
| CSS | Coarse sun sensors | 12 sensors, mounted, in groups of two, along ±X,Y and Z direction, normal to the mounting surface |
| MGMTR | Magnetometer | Mounted on the bottom U panel internally, on the +X face. |

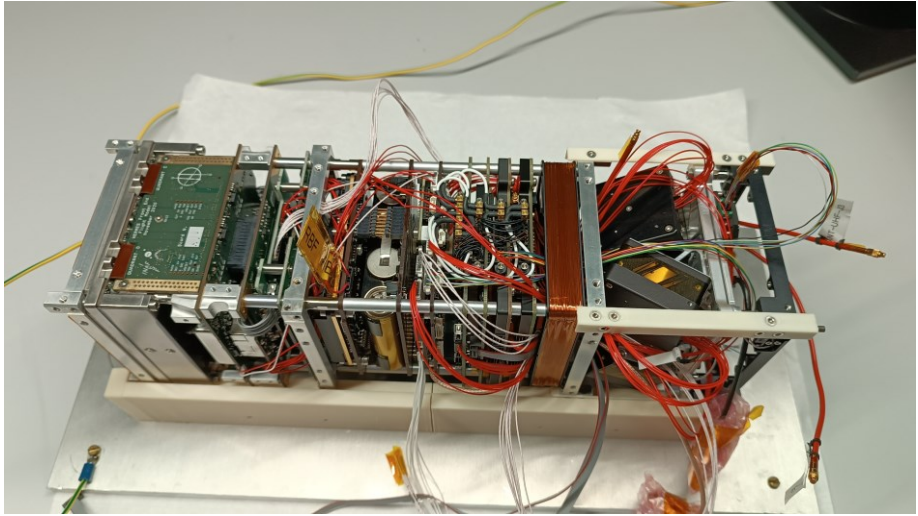

**Figure 14** HERMES H2 flight model during AIV/T activities in Politecnico di Milano clean-room

## 5. PROJECT STATUS

At the time of writing (June 2024), the manufacturing, assembly, integration and test (MAIT) program of the first two HERMES Pathfinder satellites (H1 and H2) is nearly completed. Both satellites underwent vibration tests (resonance search, random and sine vibrations) at the Politecnico di Milano shaker facility, and thermal-vacuum (TVAC) tests at Thales Alenia Space Italy.

Environmental tests verified that the HERMES Pathfinder satellites are free of defects concerning both system design and manufacturing. Figure 15 and Figure 16 show HERMES H2 completely integrated in the Politecnico of Milano clean-room and the satellite ready for the TVAC test.

MAIT program of H3-H6 satellites has started and will be completed in the next months.

## 6. CONCLUSIONS

In this paper we presented the scientific case, development and programmatic status of the HERMES Pathfinder mission, devoted to probe the temporal emission of bright high-energy transients such as Gamma-Ray Bursts (GRBs), ensuring a fast transient localization (with arcmin-level accuracy) in a field of view of several steradians exploiting the triangulation technique. HERMES Pathfinder constellation, made of 6 CubeSats 3U, will be launched in early 2025 and will operate in conjunction with the SpIRIT 6U CubeSat, launched in December 2023 and currently in a commissioning phase. At the time of writing, two out of six HERMES Pathfinder satellites have been integrated and underwent environmental testing, MAIT of the other four satellites is in progress.

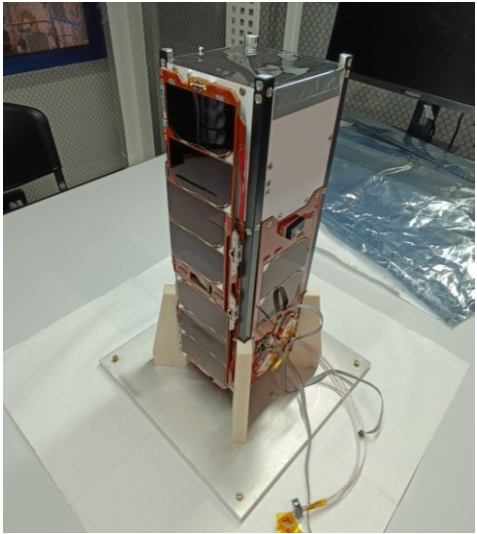 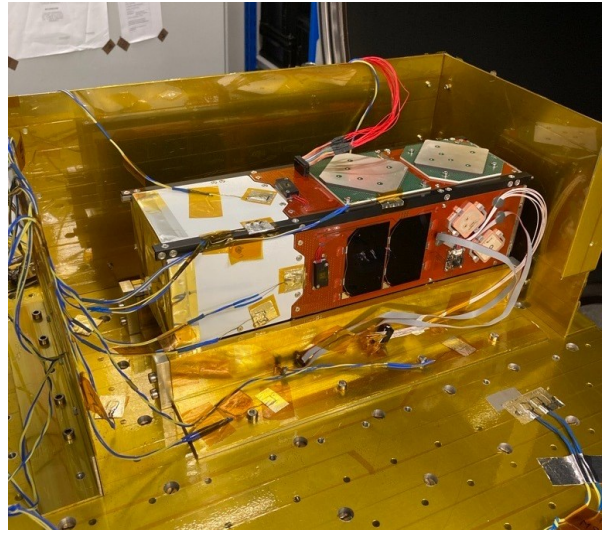

**Figure 15** HERMES Pathfinder H2 flight model completely integrated

**Figure 16** HERMES Pathfinder H2 flight model ready for TVAC environmental tests

## ACKNOWLEDGMENTS

This research acknowledges support from the European Union Horizon 2020 Research and Innovation Framework Programme under grant agreements HERMES-SP n. 821896 and AHEAD2020 n. 871158, and by ASI INAF Accordo Attuativo n. 2018-10-HH.1.2020 HERMES—Technologic Pathfinder Attività scientifiche and 2022-25-HH.0 "HERMES Pathfinder - Operazioni e sfruttamento scientifico.